\documentclass{article}

\usepackage{amsmath,amssymb,amstext}
\usepackage{url}
\usepackage{graphicx}
\usepackage[margin=1in]{geometry}

\newcommand{\BlackBox}{\rule{1.5ex}{1.5ex}}  
\newenvironment{proof}{\par\noindent{\bf Proof\ }}{\hfill\BlackBox\\[2mm]}

\newcommand {\sqb}[1]{\left[#1\right]}

\newcommand {\br}[1]{\left(#1\right)}
\newcommand {\nmFro}[1]{\Arrowvert\, #1 \,\Arrowvert_{Fro}}
\newcommand {\inpH}[2]{\left\langle #1 \right\rangle_{#2}}
\newcommand {\inpFro}[1]{\left\langle #1 \right\rangle_{Fro}}
\newcommand {\abs}[1]{\left\vert\, #1 \,\right\vert}

\newcommand{\RR}{\mathbb{R}}    
\newcommand{\Ical}{\mathcal{I}} 
\newcommand{\Ncal}{\mathcal{N}} 
\newcommand{\Tcal}{\mathcal{T}}

\newtheorem{lemma}{Lemma}

\title{Metric learning pairwise kernel for graph inference}

\author{Jean-Philippe Vert\\
Center for Computational Biology\\
Ecole des Mines de Paris\\
Fontainebleau, France\\
\texttt{Jean-Philippe.Vert@ensmp.fr}
\and
Jian Qiu\\
Department of Genome Sciences\\
University of Washington\\
Seattle, WA, USA
\and
William Stafford Noble\\
Department of Genome Sciences\\
Department of Computer Science and Engineering\\
University of Washington\\
Seattle, WA, USA}

\begin{document}

\maketitle

\begin{abstract}
Much recent work in bioinformatics has focused on the inference of
various types of biological networks, representing gene regulation,
metabolic processes, protein-protein interactions, etc.  A common
setting involves inferring network edges in a supervised fashion from
a set of high-confidence edges, possibly characterized by multiple,
heterogeneous data sets (protein sequence, gene expression, etc.).
Here, we distinguish between two modes of inference in this setting:
direct inference based upon similarities between nodes joined by an
edge, and indirect inference based upon similarities between one pair
of nodes and another pair of nodes. We propose a supervised approach
for the direct case by translating it into a distance metric learning
problem.  A relaxation of the resulting convex optimization problem
leads to the support vector machine (SVM) algorithm with a particular
kernel for pairs, which we call the {\em metric learning pairwise
kernel}.  We demonstrate, using several real biological networks, that
this direct approach often improves upon the state-of-the-art SVM for
indirect inference with the tensor product pairwise kernel.
\end{abstract}

\section{Introduction}

Increasingly, molecular and systems biology is concerned with
describing various types of subcellular networks.  These include
protein-protein interaction networks, metabolic networks, gene
regulatory and signaling pathways, and genetic interaction networks.
While some of these networks can be partly deciphered by
high-throughput experimental methods, fully constructing any such
network requires lengthy biochemical validation.  Therefore, the
automatic prediction of edges from other available data, such as
protein sequences, global network topology or gene expression
profiles, is of importance, either to speed up the elucidation of
important pathways or to complement high-throughput methods that are
subject to high levels of noise \cite{vonmering:comparative}.

Edges in a network can be inferred from relevant data in at least two
complementary ways.  For concreteness, consider a network of
protein-protein interactions derived from some noisy, high-throughput
technology.  Our confidence in the correctness of a particular edge
$A$---$B$ in this network increases if we observe, for example, that
the two proteins $A$ and $B$ localize to the same cellular compartment
or share similar evolutionary patterns \cite{ramani:exploiting,
pazos:silico, marcotte:detecting}.  Generally, in this type of {\em
direct inference}, two genes or proteins are predicted to interact if
they bear some direct similarity \emph{to each other} in the available
data.

An alternative mode of inference, which we call {\em indirect
inference}, relies upon similarities between pairs of genes or
proteins.  In the example above, our confidence in $A$---$B$ increases
if we find some other, high-confidence edge $C$---$D$ such that the
pair $\{A,B\}$ resembles $\{C,D\}$ in some meaningful fashion. Note
that in this model, the two connected proteins $A$ and $B$ might not
be similar to one another.  For example, if the goal is to detect
edges in a regulatory network by using time series expression data,
one would expect the time series of the regulated protein to be
delayed in time compared to that of the regulatory protein.
Therefore, in this case, the learning phase would involve learning
this feature from other pairs of regulatory/regulated proteins.  The
most common application of the indirect inference approach in the case
of protein-protein interaction involves comparing the amino acid
sequences of $A$ and $B$ versus $C$ and $D$ (e.g.,
\cite{sprinzak:correlated, gomez:learning, martin:predicting,
ben-hur:kernel}).

Indirect inference amounts to a straightforward application of the
machine learning paradigm to the problem of edge inference: each edge
is an example, and the task is to learn by example to discriminate
between ``true'' and ``false'' edges.  Not surprisingly, therefore,
several machine learning algorithms have been applied to predict
network edges from properties of protein pairs.  For example, in the
context of machine learning with support vector machines (SVM) and
kernel methods, Ben-Hur and Noble \cite{ben-hur:kernel} describe how
to map an embedding of individual proteins onto an embedding of pairs
of proteins.  The mapping defines two pairs of proteins as similar to
each other when each protein in a pair is similar to one corresponding
protein in the other pair. In practice, the mapping is defined by
deriving a kernel function on pairs of proteins from a kernel function
$K_g$ on individual proteins, obtained by a tensorization of the
initial feature space.  We therefore call this pairwise kernel, shown
below, the \emph{tensor product pairwise kernel} (TPPK):
\begin{equation}\label{eq:ttpk}
K_{TPPK}\br{\br{x_{1},x_{2}},\br{x_{3},x_{4}}}
=  K_{g}\br{x_{1},x_{3}}  K_{g}\br{x_{2},x_{4}} +  K_{g}\br{x_{1},x_{4}} K_{g}\br{x_{2},x_{3}}     \,.
\end{equation}

Less attention has been paid to the use of machine learning approaches
in the direct inference paradigm.  Two exceptions are the works of
Yamanishi {\em et al.} \cite{yamanishi:protein} and Vert {\em et al.}
\cite{vert:supervised}, who derive supervised machine learning
algorithms to optimize the measure of similarity that underlies the
direct approach by learning from examples of interacting and
non-interacting pairs.  Yamanishi {\em et al.} employ kernel canonical
correlation analysis to embed the proteins into a feature space where
distances are expected to correlate with the presence or absence of
interactions between protein pairs.  Vert {\em et al.}  highlight the
similarity of this approach with the problem of distance metric
learning \cite{Xing2003Distance}, while proposing an algorithm for
that purpose.

Both of these direct inference approaches, however, suffer from two
important drawbacks.  First, they are based on the optimization of a
proxy function that is slightly different from the objective of the
embedding, namely, finding a distance metric such that
interacting/non-interacting pairs fall above/below some threshold.
Second, the methods of \cite{yamanishi:protein} and
\cite{vert:supervised} are applicable only when the known part of the
network used for training is defined by a subset of proteins in the
network.  In other words, in order to apply these methods, we must
have a complete set of high-confidence edges for one set of proteins,
from which we can infer edges in the rest of the network.  This
setting is unrealistic.  In practice, our training data will generally
consist of known positive and negative edges distributed throughout
the target network.

In this paper we propose a convex formulation for supervised learning
in the direct inference paradigm that overcomes both of the
limitations mentioned above. We show that a slight relaxation of this
formulation bears surprising similarities with the supervised approach
of \cite{ben-hur:kernel}, in the sense that it amounts to defining a
kernel between pairs of proteins from a kernel between individual
proteins.  We therefore call our method the \emph{metric learning
pairwise kernel} (MLPK).  An important property of this formulation as
an SVM is the possibility to learn from several data types
simultaneously by combining kernels, which is of particular importance
in various bioinformatics applications \cite{pavlidis:gene,
lanckriet:statistical}.

We validate the MLPK approach on the task of reconstructing two yeast
networks: the network of metabolic pathways and the co-complex
network.  In each case, the network is inferred from a variety of
genomic and proteomic data, including protein amino acid sequences,
gene expression levels over a large set of experiments, and protein
cellular localization. We show that the MLPK approach nearly always
provides better prediction performance than the state-of-the-art TPPK
approach.

\section{Algorithm}

Let us assume that a gene is represented by a vector $x \in \RR^p$ of
genomic data such as a microarray expression profile. The problem of
supervised gene network inference is, given a set of $n$ genes
$x_{1},\ldots,x_{n}$ and a training set $\Tcal = \Ical \cup \Ncal
\subset [1,n]^2$ of interacting ($\Ical$) and non-interacting
($\Ncal$) pairs, to predict whether pairs of genes not in the training
set interact or not. Following \cite{vert:supervised}, we note that a
possible approach to solve this problem is to learn a distance metric
$d$ between genes with the property that pairs of nearby genes with
respect to $d$ are connected by an edge, while pairs of genes far from
each other are not. If such a metric is available, then the prediction
of an edge between a candidate pair of genes simply amounts to
computing their distance to each other and predicting an edge if the
distance is below a threshold.

\subsection{Distance metric learning}

More formally, let us investigate distance metrics obtained by linear
transformations of the input space. Such metrics are indexed by
symmetric positive semidefinite matrices $M$ as follows:
$$
d_{M}(x,x') = \br{x-x'}^\top M\br{x-x'} \,.
$$ Our goal is to learn a distance metric which separates interacting
from non-interacting pairs, while controlling over-fitting to the
training set. Following the spirit of the SVM algorithm, we enforce an
arbitrary margin of $1$ between the distances of interacting and
non-interacting variables up to slack variables, and control the
Frobenius norm of $M$ by considering the following problem:
\begin{equation}\label{eq:opt1}
\min_{\gamma, M,\zeta} \nmFro{M}^2 + C\sum_{(i,j) \in \Tcal} \zeta_{ij},
\end{equation}
under the constraints:
\begin{equation}\label{eq:opt2}
\begin{split}
\zeta_{ij} &\geq 0\,,\qquad(i,j) \in \Tcal\,,\\
d_{M}(x_{i},x_{j}) &\leq \gamma-1+\zeta_{ij}, \quad (i,j) \in \Ical\,,\\
d_{M}(x_{i},x_{j}) &\geq \gamma+1-\zeta_{ij}, \quad (i,j) \in \Ncal\,,\\
M &\succeq 0\,.
\end{split}
\end{equation}
In order to solve this problem we first prove the following extension to the representer theorem \cite{Kimeldorf1971Some}:
\begin{lemma}\label{lem:representer}
The solution of (\ref{eq:opt1}-\ref{eq:opt2}) can be expanded as:
$$
M = \sum_{(i,j)\in\Tcal} \alpha_{ij}(x_{i}-x_{j})(x_{i}-x_{j})^\top\;,
$$
with $\alpha_{ij}\in\RR$ for $(i,j)\in\Tcal$.
\end{lemma}
\begin{proof}
For any pair $(i,j)$, let us denote $u_{ij}=x_i-x_j$, and let $D_{ij}$
be the $p\times p$ matrix $D_{ij} = (x_{i}-x_{j})(x_{i}-x_{j})^\top =
u_{ij} u_{ij}^\top$. Then we can rewrite
$$
d_{M}(x_{i},x_{j}) = \inpH{M,D_{ij}}{Fro}\,,
$$
where $\inpH{A,B}{Fro} = Trace(A^\top B)$ is the Frobenius inner
product. Introducing the hinge loss function $L(y,y')=\max(1-yy',0)$
for $y,y'\in\RR$, we can eliminate the slack variables and rewrite the
problem (\ref{eq:opt1}-\ref{eq:opt2}) as:
\begin{equation}\label{eq:hinge}
\min_{M \succeq 0 , \gamma\in\RR}   \nmFro{M}^2 + C \sum_{(i,j) \in \Tcal} L(\inpH{M,D_{ij}}{Fro} - \gamma , y_i)\;.
\end{equation}
This shows that the optimization problem is in fact equivalent, up to
the positive semidefinitiveness constraint, to an SVM in the linear
space of symmetric matrices endowed with the Frobenius inner
product. Each edge example is then mapped to the matrix $D_{ij}$. In
particular, if the constraint on $M$ was not present, then Lemma
\ref{lem:representer} would be exactly the representer theorem. Here
we need to show that it still holds with the constraint $M\succeq
0$. For this purpose let $M\succeq 0$ and $\gamma\in\RR$ be the
solution of (\ref{eq:opt1}-\ref{eq:opt2}). $M$ can be uniquely
decomposed as $M=M_S + M_\perp$, where $M_S$ is in the linear span of
$\br{D_{ij} , (i,j)\in\Tcal}$ and $\inpFro{M_\perp , D_{ij}}=0$ for
$(i,j)\in\Tcal$. By the Pythagorean theorem we have $\nmFro{M}^2 =
\nmFro{M_S}^2 + \nmFro{M_\perp}^2$, so if $M_\perp \neq 0$ the
functional minimized in (\ref{eq:hinge}) is strictly smaller at
$\br{M_S,\gamma}$ than at $\br{M,\gamma}$; this would be a
contradiction if $M_S\succeq 0$. Therefore, to prove the lemma it
suffices to show $M_S \succeq 0$.  Let $v\in\RR^p$ be any vector. We
can decompose that vector uniquely as $v=v_S+v_\perp$, where $v_S$ is
in the linear span of the $u_{ij} , \br{i,j}\in\Tcal$ and
$v_\perp^\top u_{ij}=0$ for $(i,j)\in\Tcal$. We then have $M_S v_\perp
= 0$ and $M_\perp v_S=0$, and therefore
$$
v^\top M_S v = v_S^\top M_S v_S = v_S^\top M_S v_S + v_S^\top M_\perp v_S = v_S^\top M v_S \geq 0\;,
$$ where we used the fact that $M\succeq 0$ in the last
inequality. This is true for any $v\in\RR^p$, which shows that
$M_S\succeq 0$, concluding the proof.
\end{proof}

\subsection{Kernelization}

By plugging the result of Lemma \ref{lem:representer} into
(\ref{eq:opt1}-\ref{eq:opt2}) we see that this problem is equivalent
to that of finding $\alpha_{ij} , (i,j)\in\Tcal$ and $\gamma$. In
order to write out the problem explicitly, let us introduce the
following kernel between two pairs $\br{x_{1},x_{2}}$ and
$\br{x_{3},x_{4}}$ :
\begin{equation}\label{eq:mlpk}
\begin{split}
K_{MLPK}\br{\br{x_{1},x_{2}},\br{x_{3},x_{4}}} & = \inpH{D_{x_{1}x_{2}},D_{x_{3}x_{4}}}{Fro}\\
&= Trace\br{\br{x_{1}-x_{2}}\br{x_{1}-x_{2}}^\top\br{x_{3}-x_{4}}\br{x_{3}-x_{4}}^\top}\\
&= \br{\br{x_{1}-x_{2}}^\top \br{x_{3}-x_{4}}}^2\\
&= \br{x_{1}^\top x_{3} -x_{1}^\top x_{4}-x_{2}^\top x_{3} + x_{2}^\top x_{4}}^2\,.
\end{split}
\end{equation}
This kernel is positive definite because it is the Frobenius inner
product between the matrices $D_{ab}$ representing the
pairs. Moreover, although $K_{MLPK}$ is formally defined for ordered
pairs only, we observe that it is invariant by permutation of the
elements of each pair (e.g., when $x_1$ and $x_2$ are flipped). It can
therefore be considered as a positive definite kernel over the set of
\emph{unordered pairs}, seen as the quotient space of the set of
ordered proteins with respect to the equivalence relation of
permutation among each pair. We call this kernel for unordered pairs
the \emph{metric learning pairwise kernel} (MLPK), hence the notation
$K_{MLPK}$.

In order to express the problem (\ref{eq:opt1}-\ref{eq:opt2}) in terms
of the $\alpha$ variables provided by Lemma \ref{lem:representer}, we
need to express the constraint $M\succeq 0$ in terms of
$\alpha$. Denoting pairs of indices $t=(i,j)$, Lemma
\ref{lem:representer} ensures that $M$ can be written as
$M=\sum_{t\in\Tcal} \alpha_t u_t u_t^\top$. As we showed in the proof
of Lemma \ref{lem:representer}, this implies that $M$ is null on the
space orthogonal to the linear span of $\br{u_t,t\in\Tcal}$.
Therefore, $M\succeq 0$ if and only if $v^\top M v \geq 0$ for any $v$
in the linear span of $\br{u_t,t\in\Tcal}$. This is equivalent to the
fact that the $\abs{\Tcal}\times\abs{\Tcal}$ matrix $F$ defined by
$F_{t,t'}=u_t^\top M u_{t'}$ is positive semidefinite. Finally, if we
denote by $F_t$ the $\abs{\Tcal}\times\abs{\Tcal}$ matrix whose
$(t_1,t_2)$ entry is $u_{t_1}^\top D_t u_{t_2} = u_{t_1}^\top u_t
u_t^\top u_{t_2}$, this is equivalent to $\sum_{t\in\Tcal} \alpha_t
F_t \succeq 0$.

Plugging the representation of Lemma \ref{lem:representer} into
(\ref{eq:opt1}-\ref{eq:opt2}), and replacing the Frobenius inner
product by the MLPK kernel, we show that the problem is equivalent
to
\begin{equation}\label{eq:opt1ker}
\min_{\alpha,\gamma,\zeta} \sum_{(i,j)\in\Tcal}\sum_{(k,l)\Tcal} \alpha_{ij}\alpha_{kl}K_{MLPK}\br{(x_i,x_j),(x_k,x_l)} + C\sum_{(i,j) \in \Tcal} \zeta_{ij},
\end{equation}
under the constraints:
\begin{equation}\label{eq:opt2ker}
\begin{split}
\zeta_{ij} &\geq 0\,,\qquad(i,j) \in \Tcal\,,\\
\sum_{(k,l)\in\Tcal} \alpha_{kl} K_{MLPK}\br{(x_i,x_j),(x_k,x_l)} &\leq \gamma-1+\zeta_{ij}, \quad (i,j) \in \Ical\,,\\
\sum_{(k,l)\in\Tcal} \alpha_{kl} K_{MLPK}\br{(x_i,x_j),(x_k,x_l)} &\geq \gamma+1-\zeta_{ij}, \quad (i,j) \in \Ncal\,,\\
\sum_{(k,l)\in\Tcal} \alpha_{kl} F_{kl} & \succeq 0
\end{split}
\end{equation}

An important property of this problem is that the data only appear
through the kernel $K_{MLPK}$ and the matrices $F_{ij}$. Furthermore,
the MLPK kernel itself (\ref{eq:mlpk}) computed between two pairs of
vectors only involves inner products between the vectors; similarly
the $(t_1,t_2)$-th entry of the matrix $F_{t}$ is a product of inner
products, which can easily be computed from the inner products of the
data themselves. As a result, we can apply the kernel trick to extend
the problem (\ref{eq:opt1ker}-\ref{eq:opt2ker}) to any data space
endowed with a positive definite kernel $K_g$.  The resulting MLPK
kernel between pairs becomes
$$
K_{MLPK}\br{\br{x_{1},x_{2}},\br{x_{3},x_{4}}}
= \br{ K_{g}\br{x_{1},x_{3}} - K_{g}\br{x_{1},x_{4}} - K_{g}\br{x_{2},x_{3}} + K_{g}\br{x_{2},x_{4}}  }^2 \,,
$$
and for any three pairs $t=(i,j),t_1=(i_1,j_1),t_2=(i_2,j_2)$ in
$\Tcal$ the entry $(t_1,t_2)$ of $F_t$ is
\begin{multline*}
\sqb{K_g\br{x_{i_1},x_i} - K_g\br{x_{i_1},x_j} - K_g\br{x_{j_1},x_i} + K_g\br{x_{j_1},x_j}}\\
\times \sqb{K_g\br{x_{i_2},x_i} - K_g\br{x_{i_2},x_j} - K_g\br{x_{j_2},x_i} + K_g\br{x_{j_2},x_j}}\;.
\end{multline*}

\subsection{Relaxation}

The problem (\ref{eq:opt1ker}-\ref{eq:opt2ker}) is a convex problem
over the cone of positive semidefinite matrices that can in theory be
solved by algorithms such as interior-point methods
\cite{boyd:convex}.  The dimension of this problem, however, is
$2\abs{\Tcal}+1$. This is typically of the order of several thousands
for small biological networks with a few hundreds or thousands
vertices, which poses serious convergence issues for general-purpose
optimization software.

If we relax the condition $M\succeq 0$ in the original problem, then
it becomes the quadratic program of the SVM, for which dedicated
optimization algorithms have been developed: current implementations
of SVM easily handle several tens of thousands of dimensions
\cite{scholkopf:learning}. The obvious drawback of this relaxation is
that if the matrix $M$ is not positive semidefinite, then it does not
define a metric. Although this can be a serious problem for classical
applications of distance metric learning such as clustering
\cite{Xing2003Distance}, we note that in our case the goal of metric
learning is just to provide a decision function $f(x,x') =
d_{M}(x,x')$ for predicting connected pairs, and negativity of this
decision function is not a problem in itself.  Therefore, we propose
to relax the constraint $M\succeq 0$, or equivalently
$\sum_{(kl)\in\Tcal} \alpha_{kl} F_{k,l} \succeq 0$ in
(\ref{eq:opt2ker}), and to solve the initial problem using an SVM over
pairs with the MLPK kernel (\ref{eq:mlpk}).

\section{Experiments}

We present below a comparison of the previously described TPPK kernel
and the new MLPK kernel for the reconstruction of two biological
networks: the metabolic network and the co-complex protein
network. For each network, we cast the problem of network
reconstruction as a binary classification problem, where the presence
or absence of edges must be inferred from various types of data
relevant to the proteins.  Because the network contains relatively few
edges compared to the total number of possible pairs, we created a
balanced dataset by keeping all known edges as positive examples and
randomly sampling an equal number of absent edges as negative
examples.  We compare the utilities of the TPPK and MLPK kernels in
this context by assessing the performance of an SVM for edge
prediction in a five-fold cross-validation experiment repeated three
times (3x5cv) with different random folds. At each fold, the
regularization parameter $C$ of the SVM is chosen among $18$ values
evenly log-spaced on the interval $[10^{-4},50]$ by minimizing the
classification error estimated by five-fold cross-validation within
the training set only.

\subsection{Metabolic network}

Most biochemical reactions in living organisms are catalyzed by
particular proteins called enzymes, and occur sequentially to form
metabolic pathways. For example, the degradation of glucose into
pyruvate (called glycolysis) involves a sequence of ten chemical
reactions catalyzed by ten enzymes. The metabolic gene network is
defined as an undirected graph with enzymes as vertices and with edges
connecting pairs of enzymes that can catalyze successive chemical
reactions.  The reconstruction of metabolic pathways for various
organisms is of critical importance, e.g., to find new ways to
synthesize chemical compounds of interest.  This problem motivated
earlier work on supervised graph inference \cite{yamanishi:protein,
vert:supervised}.  Focusing on the budding yeast \emph{S. cerevisiae},
we collected the metabolic network and genomic data used in
\cite{yamanishi:protein}. The network was extracted from the KEGG
database and contains 769 vertices and 3702 undirected edges.

In order to infer the network, various independent data about the
proteins can be used. In this experiment, we use four relevant sources
of data provided by \cite{yamanishi:protein}: (1) a set of 157 gene
expression measurements obtained from DNA microarrays; (2) the
phylogenetic profiles of the genes, represented as 145-bit vectors
indicating the presence or absence of each gene in 145 fully sequenced
genomes; (3) the protein's localization in the cell determined
experimentally \cite{huh:global}, represented as 23-bit vectors
corresponding to 23 cellular compartments, and (4) yeast two-hybrid
protein-protein interaction data \cite{vonmering:comparative},
represented as a network.  For the first three data sets, a Gaussian
RBF kernel was used to represent the data as a kernel matrix.  For the
yeast two-hybrid network,we use a diffusion kernel
\cite{kondor:diffusion}.  Additionally, we considered a fifth kernel
obtained by summing the first four kernels. This is a simple approach
to data integration that has proved useful in previous work
\cite{pavlidis:gene, yamanishi:extraction}.

\begin{table}
\caption{{\bf Performance on reconstruction of the yeast metabolic and
  co-complex networks.} The table lists, for each network and each
  type of data, the accuracy and area under the ROC curve obtained by
  each pairwise kernel.  Values in the tables are means and standard
  errors in a 3x5cv experiment.  TPPK is the tensor product pairwise
  kernel, and MLPK is the metric learning pairwise kernel.}
\begin{center}
\begin{tabular}{c|c|c|c|c|c}
& & \multicolumn{2}{|c|}{MLPK} & \multicolumn{2}{|c}{TPPK} \\
Network & Data & Accuracy & AUC & Accuracy & AUC \\
\hline
\hline
& Expression & $77.8 \pm 0.2$ & $84.5 \pm 0.1$ & $76.7 \pm 0.3$ & $83.3 \pm 0.2$ \\
& Localization & $63.9 \pm 0.4$ & $68.2 \pm 0.4$ & $62.3 \pm 0.1$ & $65.8 \pm 0.4$ \\
Metabolic & Phylogenetic profile & $79.8 \pm 0.1$ & $84.9 \pm 0.2$ & $78.4 \pm 0.1$ & $83.4 \pm 0.4$ \\
& Yeast two-hybrid & $76.6 \pm 0.2$ & $82.0 \pm 0.1$ & $59.2 \pm 0.1$ & $65.1 \pm 0.6$ \\
& Sum & $83.9 \pm 0.4$ & $90.9 \pm 0.3$ & $84.2 \pm 0.5$ & $91.1 \pm 0.3$ \\
\hline
\hline
& Localization & $76.5 \pm 0.1$ & $76.8 \pm 0.1$ & $ 79.6 \pm 0.1 $ & $ 83.1 \pm 0.1 $ \\
Co-complex & Chip-chip & $82.4 \pm 0.3$ & $89.7 \pm 0.1$ & $63.8 \pm 0.1$ & $67.9 \pm 0.3$ \\
& Pfam & $92.2 \pm 0.2$ & $98.2 \pm 0.1$ & $85.5 \pm 0.1$ & $91.7 \pm 0.2$ \\
& PSI-BLAST & $90.0 \pm 0.3$ & $97.3 \pm 0.1$ & $88.3 \pm 0.1$ & $93.6 \pm 0.2$ \\
\end{tabular}
\end{center}
\label{table:results}
\end{table}

Table \ref{table:results} (top) shows the performance of each pairwise
kernel for the five data sets. The MLPK is never worse than the TPPK
kernel. The two kernels have similar performance on the sum kernel;
MLPK is slightly better than TPPK on the expression, localization and
phylogenetic profile kernels, and much better on the yeast two-hybrid
dataset ($76.6$\% vs. $59.2$\% in accuracy).

Interestingly, we note that although connected pairs, i.e., pairs of
enzymes acting successively in a pathway, are expected to have similar
expression, phylogenetic profiles and localization (explaining the
good performance of the MLPK on these datasets), the indirect approach
implemented by the TPPK also gives good results for these data.  This
result implies that for these data, interacting pairs in the training
set are often similar not only to each other but also to other
interacting pairs in the training set.  This observation is not
surprising because, for example, if two proteins of the test set are
co-localized in a particular organelle, it is likely that interacting
pairs of proteins co-localized in the same organelle are also present
in the training set, because there are not so many organelles where
connected proteins can be.

In the case of yeast two-hybrid data, on the other hand, the kernel
between single proteins is defined as a diffusion kernel over the
yeast two-hybrid graph. One can speculate that, in that case,
similarity between pairs can be easily assessed and used by the MLPK
to predict edges, but similarity between pairs as defined by the TPPK
kernel is less likely to be observed.  In a sense, the dimensionality
of the feature space of the diffusion kernels is much larger than that
defined by the other kernels, and a protein is only close to its
neighbors in the yeast two-hybrid graph.

\subsection{Protein complex network}

Many proteins carry out their biological functions by acting together
in multi-protein structures known as complexes.  Understanding protein
function therefore requires identification of these complexes.  In the
co-complex network, nodes are proteins, and an edge between proteins
$A$ and $B$ exists if $A$ and $B$ are members of the same protein
complex.  Some high-throughput experimental methods, such as tandem
affinity purification followed by mass spectrometry, explicitly
identify these co-complex relationships, albeit in a noisy fashion.
Also, computational methods exist for inferring the co-complex network
from individual data types or from multiple data types simultaneously
\cite{jansen:bayesian, qi:evaluation}. We derived the co-complex data
set based on an intersection of the manually curated MIPS complex
catalogue \cite{mewes:mips} and the BIND complex data set
\cite{bader:bind}.  The co-complex network contains 3280 edges
connecting 797 proteins.  In addition, our data set contains 3081
proteins with no co-complex relationships.

For this evaluation, we again use four different data sets that we
consider relevant to the co-complex network.  The first data set is
the same localization data that we used above \cite{huh:global}.  The
second is derived from a chip-based version of the chromatin
immunoprecipitation assay (so-called ``ChIP-chip'' data)
\cite{harbison:transcriptional}.  This assay provides evidence that a
transcription factor binds to the upstream region of a given gene and
is likely to regulate the expression of the given gene.  Our data set
contains data for 113 transcription factors, and so yields a vector of
length 113 for each protein.  The final two data sets are derived from
the amino acid sequences of the yeast proteins.  For the first, we
compared each yeast protein to every model in the Pfam database of
protein domain HMMs (\url{pfam.wustl.edu}) and recorded the E-value of
the match.  This comparison yields a vector of length 8183 for each
protein.  Finally, in a similar fashion, we compared each yeast
protein to each protein in the Swiss-Prot database version 40
(\url{ca.expasy.org/sprot}) using PSI-BLAST \cite{altschul:gapped},
yielding vectors of length 101,602.  Each of the four data sets is
represented using a scalar product kernel.

We used the same experimental procedure to compare the quality of edge
predictors for the co-complex network using MLPK and TPPK.  The
results, shown in Table~\ref{table:results} (bottom), again show the
value of the MLPK approach.  Using either performance metric (accuracy
or ROC area), the MLPK approach performs better than the TPPK approach
on three out of four data sets.

Most striking is the improvement for the ChIP-chip data set (accuracy
from $63.8$\% to $82.4$\%).  This result is expected, because we know
that proteins in the same complex must act in concert.  As such, they
are typically regulated by a common set of transcription factors.

In contrast, the MLPK approach does not fare better than TPPK on the
localization data set.  This is, at first, suprising because two
proteins must co-localize in order to participate in a common complex.
This problem is thus an example of the direct inference case for which
the MLPK is designed.  However, the localization data is somewhat
complex because (1) only approximately 70\% of yeast proteins are
assigned any localization at all, and (2) many proteins are assigned
to multiple locations.  As a result, among 3280 positive edges in the
training set, only 1852 (56\%) of those protein pairs share exactly
the same localization.  Furthermore, 550 (16.8\%) of the 3280 negative
edges used in training connect proteins with the same localization,
primarily ``Unknown.''  These factors make direct inference using this
data set difficult.  The indirect method, by contrast, is apparently
able to identify useful relationships, corresponding to specific
localizations, that are enriched among the positive pairs relative to
the negative pairs.

\section{Discussion}

We showed how a particular formulation of metric distance learning for
graph inference can be formulated as a convex optimization problem and
can be applied to any data set endowed with a positive definite
kernel. A relaxation of this problem leads to the SVM algorithm with
the new MLPK kernel (\ref{eq:mlpk}) between pairs. Experiments on two
biological networks confirm the value of this approach for the
reconstruction of biological network from heterogeneous genomic and
proteomic data.

Beyond the direct and indirect approaches to graph inference mentioned
in the introduction, there exist many alternative ways to infer
networks, such as estimating conditional independence between vertices
with Bayesian networks \cite{Friedman2000Using}.  An interesting
property of methods based on supervised learning, such as the SVM with
the TPPK and MLPK kernels, is the limited hypothesis made on the
nature of the edges; the only hypothesis made is that there is
information related to the presence or absence of edges in the data,
and we let the learning algorithm model this information. The good
accuracy obtained on two completely different networks (metabolic and
co-complex) supports the general utility of this approach.

An interesting and important avenue for future research is the
extension of these approaches to inference of directed graphs, e.g.,
regulatory networks. Although the TPPK and MLPK approaches are not
adapted as such to this problem, variants involving for example
kernels between ordered pairs could be studied.

\section{Acknowledgements}
This work was funded by NIH award R33~HG003070.

\bibliographystyle{plain}

\begin{thebibliography}{10}

\bibitem{altschul:gapped}
S.~F. Altschul, T.~L. Madden, A.~A. Schaffer, J.~Zhang, Z.~Zhang, W.~Miller,
  and D.~J. Lipman.
\newblock Gapped {BLAST} and {PSI-BLAST}: {A} new generation of protein
  database search programs.
\newblock {\em Nucleic Acids Research}, 25:3389--3402, 1997.

\bibitem{bader:bind}
G.~D. Bader, I.~Donaldson, C.~Wolting, B.~F. Ouellette, T.~Pawson, and C.~W.
  Hogue.
\newblock {BIND}--the biomolecular interaction network database.
\newblock {\em Nucleic Acids Res}, 29(1):242--245, 2001.

\bibitem{ben-hur:kernel}
A.~Ben-Hur and W.~S. Noble.
\newblock Kernel methods for predicting protein-protein interactions.
\newblock {\em Bioinformatics}, 21 suppl 1:i38--i46, 2005.

\bibitem{boyd:convex}
S.~Boyd and L.~Vandenberghe.
\newblock {\em Convex Optimization}.
\newblock Prentice-{H}all, 2003.
\newblock To appear. Available at
  http://www.stanford.edu/\~{}boyd/cvxbook.html.

\bibitem{Friedman2000Using}
N.~Friedman, M.~Linial, I.~Nachman, and D.~Pe'er.
\newblock Using {B}ayesian {N}etworks to {A}nalyze {E}xpression {D}ata.
\newblock {\em J. {C}omput. {B}iol.}, 7(3-4):601--620, 2000.

\bibitem{gomez:learning}
S.~M. Gomez, W.~S. Noble, and A.~Rzhetsky.
\newblock Learning to predict protein-protein interactions.
\newblock {\em Bioinformatics}, 19:1875--1881, 2003.

\bibitem{harbison:transcriptional}
C.T. Harbison, D.B. Gordon, T.I. Lee, N.~Rinaldi, K.D. Macisaac, T.D. Danford,
  N.M. Hannett, J.-B. Tagne, D.B. Reynolds, J.~Yoo, E.G. Jennings,
  J.~Zeitlinger, D.K. Pokholok, M.~Kellis, P.A. Rolfe, K.T. Takusagawa, E.S.
  Lander, D.K. Gifford, E.~Fraenkel, and R.A. Young.
\newblock Transcriptional regulatory code of a eukaryotic genome.
\newblock {\em Nature}, 431:99--104, 2004.

\bibitem{huh:global}
W.~K. Huh, J.~V. Falvo, L.~C. Gerke, A.~S. Carroll, R.~W. Howson, J.~S.
  Weissman, and E.~K. O'Shea.
\newblock Global analysis of protein localization in budding yeast.
\newblock {\em Nature}, 425:686--691, October 16 2003.

\bibitem{jansen:bayesian}
R.~Jansen, H.~Yu, D.~Greenbaum, Y.~Kluger, N.~J. Krogan, S.~Chung, A.~Emili,
  M.~Snyder, J.~F. Greenblatt, and M.~Gerstein.
\newblock A {Bayesian} networks approach for predicting protein-protein
  interactions from genomic data.
\newblock {\em Science}, 302:449--453, 2003.

\bibitem{Kimeldorf1971Some}
G.~S. Kimeldorf and G.~Wahba.
\newblock Some results on {T}chebycheffian spline functions.
\newblock {\em J. {M}ath. {A}nal. {A}ppl.}, 33:82--95, 1971.

\bibitem{kondor:diffusion}
R.~I. Kondor and J.~Lafferty.
\newblock Diffusion kernels on graphs and other discrete input spaces.
\newblock In C.~Sammut and A.~Hoffmann, editors, {\em Proceedings of the
  International Conference on Machine Learning}. Morgan Kaufmann, 2002.

\bibitem{lanckriet:statistical}
G.~R.~G. Lanckriet, T.~De Bie, N.~Cristianini, M.~I. Jordan, and W.~S. Noble.
\newblock A statistical framework for genomic data fusion.
\newblock {\em Bioinformatics}, 20(16):2626--2635, 2004.

\bibitem{marcotte:detecting}
E.~M. Marcotte, M.~Pellegrini, H.-L. Ng, D.~W. Rice, T.~O. Yeates, and
  D.~Eisenberg.
\newblock Detecting protein function and protein-protein interactions from
  genome sequences.
\newblock {\em Science}, 285:751--753, 1999.

\bibitem{martin:predicting}
S.~Martin, D.~Roe, and J-L. Faulon.
\newblock Predicting protein-protein interactions using signature products.
\newblock {\em Bioinformatics}, 21(2):218--226, 2005.

\bibitem{mewes:mips}
H.~W. Mewes, D.~Frishman, C.~Gruber, B.~Geier, D.~Haase, A.~Kaps, K.~Lemcke,
  G.~Mannhaupt, F.~Pfeiffer, C~Sch{\"{u}}ller, S.~Stocker, and B.~Weil.
\newblock {MIPS}: a database for genomes and protein sequences.
\newblock {\em Nucleic Acids Research}, 28(1):37--40, 2000.

\bibitem{pavlidis:gene}
P.~Pavlidis, J.~Weston, J.~Cai, and W.~N. Grundy.
\newblock Gene functional classification from heterogeneous data.
\newblock In {\em Proceedings of the Fifth Annual International Conference on
  Computational Molecular Biology}, pages 242--248, 2001.

\bibitem{pazos:silico}
F.~Pazos and A.~Valencia.
\newblock In silico two-hybrid system for the selection of physically
  interacting protein pairs.
\newblock {\em Proteins: Structure, Function and Genetics}, 47(2):219--227,
  2002.

\bibitem{qi:evaluation}
Y.~Qi, Z.~Bar-Joseph, and J.~Klein-Seetharaman.
\newblock Evaluation of different biological data and computational
  classification methods for use in protein interaction prediction.
\newblock {\em Proteins: Structure, Function, and Bioinformatics}, 63:490--500,
  2006.

\bibitem{ramani:exploiting}
A.K. Ramani and E.M. Marcotte.
\newblock Exploiting the co-evolution of interacting proteins to discover
  interaction specificity.
\newblock {\em Journal of Molecular Biology}, 327(1):273--284, 2003.

\bibitem{scholkopf:learning}
B.~Sch\"{o}lkopf and A.~Smola.
\newblock {\em Learning with Kernels}.
\newblock MIT Press, Cambridge, MA, 2002.

\bibitem{sprinzak:correlated}
E.~Sprinzak and H.~Margalit.
\newblock Correlated sequence-signatures as markers of protein-protein
  interaction.
\newblock {\em Journal of Molecular Biology}, 311:681--692, 2001.

\bibitem{vert:supervised}
J.-P. Vert and Y.~Yamanishi.
\newblock Supervised graph inference.
\newblock In L.~K. Saul, Y.~Weiss, and L.~Bottou, editors, {\em Advances in
  Neural Information Processing Systems 17}, pages 1433--1440. MIT Press,
  Cambridge, MA, 2005.

\bibitem{vonmering:comparative}
C.~{von Mering}, R.~Krause, B.~Snel, M.~Cornell, S.~G. Olivier, S.~Fields, and
  P.~Bork.
\newblock Comparative assessment of large-scale data sets of protein-protein
  interactions.
\newblock {\em Nature}, 417:399--403, 2002.

\bibitem{Xing2003Distance}
E.P. Xing, A.Y. Ng, M.I. Jordan, and S.~Russell.
\newblock Distance metric learning with application to clustering with
  side-information.
\newblock In S.~Thrun S.~Becker and K.~Obermayer, editors, {\em Adv. Neural.
  Inform. Process Syst.}, volume~15, pages 505--512, Cambridge, MA, 2003. MIT
  Press.

\bibitem{yamanishi:protein}
Y.~Yamanishi, J.-P. Vert, and M.~Kanehisa.
\newblock Protein network inference from multiple genomic data: a supervised
  approach.
\newblock {\em Bioinformatics}, 20:i363--i370, 2004.

\bibitem{yamanishi:extraction}
Y.~Yamanishi, J.-P. Vert, A.~Nakaya, and M.~Kanehisa.
\newblock Extraction of correlated gene clusters from multiple genomic data by
  generalized kernel canonical correlation analysis.
\newblock {\em Bioinformatics}, 2003.
\newblock To appear.

\end{thebibliography}

\end{document}